\documentclass{pasa}%

\usepackage{graphicx}	
\usepackage{amsmath}	
\usepackage{amssymb}	
\usepackage{empheq}
\graphicspath{{./}{figures/}}

\title[Impact of Tandem Calibration]{The Impact of Tandem Redundant/Sky-Based Calibration in MWA Phase II Data Analysis}

\author[Zheng Zhang et al.]{Zheng Zhang$^1$, Jonathan C. Pober$^1$, Wenyang Li$^1$, Bryna J. Hazelton$^{2,3}$, Miguel F. Morales$^2$, Cathryn M. Trott$^{4,5}$, Christopher H. Jordan$^{4,5}$, Ronniy C. Joseph$^{4,5}$, Adam Beardsley$^6$, Nichole Barry$^{5,7}$, Ruby Byrne$^2$, Steven J. Tingay$^{4,5}$, Aman Chokshi$^{5,7}$, Kenji Hasegawa$^{8}$, Daniel C. Jacobs$^{6}$, Adam Lanman$^{1}$, Jack L. B. Line$^{4,5}$, Christene Lynch$^{4,5}$, Benjamin McKinley$^{4,5}$, Daniel A. Mitchell$^{4,9}$, Steven Murray$^{4,5}$, Bart Pindor$^{5,7}$, Mahsa Rahimi$^{5,7}$, Keitaro Takahashi$^{10,12}$, Randall B. Wayth$^{4,5}$, Rachel L. Webster$^{5,7}$, Michael Wilensky$^{2}$, Shintaro Yoshiura$^{7}$ and Qian Zheng$^{11}$
\affil{$^1$Department of Physics, Brown University, Providence, RI 02912, USA}%
\affil{$^2$University of Washington, Department of Physics, Seattle, WA 98195, USA}
\affil{$^3$University of Washington, eScience Institute, Seattle, WA 98195, USA}
\affil{$^4$International Centre for Radio Astronomy Research, Curtin University, Perth, WA 6845, Australia}
\affil{$^5$ARC Centre of Excellence for All Sky Astrophysics in 3 Dimensions (ASTRO-3D)}
\affil{$^6$Arizona State University, School of Earth and Space Exploration, Tempe, AZ 85287, USA}
\affil{$^7$The University of Melbourne, School of Physics, Parkville, VIC 3010, Australia}
\affil{$^8$Graduate School of Science, Nagoya University, Japan}
\affil{$^9$CSIRO Astronomy and Space Science (CASS), PO Box 76, Epping, NSW 1710, Australia}
\affil{$^{10}$Faculty of Science, Kumamoto University, 2-39-1 Kurokami, Kumamoto 860-8555, Japan}
\affil{$^{11}$Shanghai Astronomical Observatory, China}
\affil{$^{12}$International Research Organization for Advanced Science and Technology, Kumamoto University, Kumamoto, 860-8555, Japan}
}%

\jid{PASA}
\doi{10.1017/pas.\the\year.xxx}
\jyear{\the\year}

\usepackage{aas_macros}
\usepackage{hyperref} 
\hypersetup{colorlinks,citecolor=blue,linkcolor=blue,urlcolor=blue}

\hypersetup{draft}

\begin{document}

\begin{frontmatter}
\maketitle

\begin{abstract}
Precise instrumental calibration is of crucial importance to 21-cm cosmology experiments. The Murchison Widefield Array's (MWA) Phase II compact configuration offers us opportunities for both redundant calibration and sky-based calibration algorithms; using the two in tandem is a potential approach to mitigate calibration errors caused by inaccurate sky models. The MWA Epoch of Reionization (EoR) experiment targets three patches of the sky (dubbed EoR0, EoR1, and EoR2) with deep observations. Previous work in \cite{Li_2018} and \cite{Wenyang_2019} studied the effect of tandem calibration on the EoR0 field and found that it yielded no significant improvement in the power spectrum over sky-based calibration alone.  
In this work, we apply similar techniques to the EoR1 field and find a distinct result: the improvements in the power spectrum from tandem calibration are significant.
To understand this result,
we analyze both the calibration solutions themselves and the effects on the power spectrum over three nights of EoR1 observations.  
We conclude that the presence of the bright radio galaxy Fornax A in EoR1 degrades the performance of sky-based calibration, which in turn enables redundant calibration to have a larger impact.
These results suggest that redundant calibration can indeed mitigate some level of model-incompleteness error. 
\end{abstract}

\end{frontmatter}

\section{INTRODUCTION }
\label{sec:intro}

The highly redshifted 21-cm signal of neutral hydrogen gas encodes the perturbation statistics of hyperfine states of neutral hydrogen gas, which trace the conditions of the radiation fields permeating the early universe.  Measurements of this signal are therefore a promising path for further constraints on the Epoch of Reionization and dark energy. 
Several experiments aiming to detect the 21-cm EoR signal are either complete or under way, such as the Donald C. Backer Precision Array for Probing the Epoch of Reionization (PAPER; \citealt{Parsons_2010}), the Low-Frequency Array (LOFAR; \citealt{LOFAR}), the Murchison Widefield Array (MWA; \citealt{tingay_goeke_bowman_emrich_ord_mitchell_morales_booler_crosse_wayth_et_al._2013,Bowman_2013}), the Giant Metrewave Radio Telescope (GMRT; \citealt{GMRT}), and the Hydrogen Epoch of Reionization Array (HERA; \citealt{DeBoer_2017}). 

However, the astrophysical foregrounds are four to five orders of magnitude brighter than the faint cosmological signal. But because the foregrounds are spectrally smooth, while the 21-cm signal has significant spectral structure, they can be in principle distinguished after performing a Fourier transform in frequency. The key challenges are precise calibration of the complicated frequency dependence of the instruments, accurate avoidance of radio frequency interference (RFI), and the effects of the ionospheric activity. 
Significant work in the past decade has led to a number of techniques aimed at overcoming these difficulties: data quality metrics to flag channels contaminated by RFI \citep{Wilensky_2019,Wenyang_2019,RFI_Josh,offringa_2015}, metrics to assess ionospheric activity \citep{Jordan_2017,Trott_2018}, and a variety of instrumental calibration methods (e.g. \citealt{Liu_2010,barry_2019,calibration}).


A leading paradigm of 21-cm analysis is to measure the power spectrum (PS) of the signal in two-dimensional (2D) cylindrical Fourier space (i.e. the $(k_{\perp},k_{\parallel})$ plane, with Fourier modes perpendicular and parallel to the plane of the sky, respectively\footnote{Strictly speaking, the sky is curved and Fourier modes do not provide an orthonormal basis for describing it.  However, as long as the fields-of-view are not too large, the distinction is negligible and is usually not accounted for in the literature; see \cite{liu_2016} for a rigorous treatment.}.  In the 2D $(k_{\perp},k_{\parallel})$ plane, foregrounds are confined to a wedge shaped region, called the ``foreground wedge''. The remaining region, namely the ``EoR window'', is expected to be contaminant free \citep{Pober_2014,Datta_2010,Parsons_2012b,Morales_2012,Vedantham:2011mh,Thyagarajan_2013,Trott_2012,liu2014epoch}.

For perfectly calibrated instruments, signal in the EoR window should only come from 21-cm signal emission in an uncontaminated observation. But in practice, small chromatic instrumental calibration errors introduce spectral structure in astrophysical foregrounds, which contaminates some Fourier modes in the EoR window and thus overwhelms the cosmological PS \citep{Morales_2019,Barry_2016}.

The two most popular methods of interferometric gain calibration are ``sky-based'' and ``redundant''; both are subject to systematic errors. Sky-based calibration uses deep foreground catalogs or maps to create model visibilities for calibration, which requires an accurate sky model; errors are introduced if the model does not capture the true sky brightness distribution.  
In particular, \citet{Barry_2016} identified a specific class of calibration errors where the instrument's point spread function (PSF) interacts with missing sources in the sky model to introduce chromatic errors and couple foreground power into the EoR window. 
Redundant calibration takes advantage of the fact that any given correlation should depend only on relative positions of two antennas, which reduces a large number of unknowns when calibrating an redundant array.   Redundant calibration is, to first order, sky-model independent but may be contaminated by non-redundancy introduced through beam variation and antenna position errors \citep{orosz_2019,Joseph2020}.
\citet{Joseph2018} found that the position offsets introduce a phase bias of the calibration solutions. This phase bias increases with the distance between bright radio sources and the pointing center, and with the flux density of these sources \citep{Joseph2018}.
\citet{Byrne_2019} also showed that calibration errors caused by sky-model incompleteness affect redundant calibration, since redundant calibration leaves degenerate parameters that must be constrained via reference to a sky model.

Although the principles and limitations of each calibration method are reasonably well-understood, the actual performance of these techniques will be affected by realistic constraints, such as algorithm implementations, array configuration, sky model accuracy, and data contamination from interference.
Recent analysis with the HERA telescope shows the potential for redundant calibration to provide some degree of information beyond sky-calibration, as well as its value as a data quality metric \citep{kern2020,dillon2020}.
\citet{Li_2018} used the compact array of Phase II of the MWA \citep{wayth_2018,Beardsley_2019}, which has both large numbers of redundant baselines and a relatively complete sampling of Fourier space (i.e. UV-coverage), to compare redundant calibration with sky-based algorithms in the EoR0 field (see \S\ref{sec:obs} for a description of the fields observed by the MWA EoR project). That work showed that the tandem combination of sky-based calibration and redundant calibration yielded small but non-negligible reductions in foreground contamination in certain modes of the EoR PS. However, \cite{Wenyang_2019} performed a similar analysis after adding an improved auto-correlation based technique to the sky-based calibration step, and found that the improvements brought about by redundant calibration became negligible. 
In this paper, we further investigate the performance of tandem redundant plus sky-based calibration in MWA Phase II data analysis in the EoR1 field, where we find a different result than that of \cite{Wenyang_2019}: redundant calibration has a significant positive impact on foreground contamination in the power spectrum even with the auto-correlation based technique applied to sky-based calibration (as will be shown in \S\ref{sec:PS}).  Before presenting the power spectrum results, we compare the calibration solutions derived from both sky and redundant calibration on a two-minute cadence from three nights of EoR1 and one night of EoR0 to investigate differences in the solutions themselves.
As in \citet{Wenyang_2019}, we use packages Fast Holographic Deconvolution (FHD) and {\tt\string OMNICAL} for state-of-the-art implementations of sky-based and redundant calibration, respectively. 

The organization of this paper is as follows.
In Section~\ref{sec:obs}, we describe the compact array of MWA Phase II and the observations analysed in this work. In Section~\ref{sec:cal}, we review the methodology and mathematical framework of sky-based calibration and redundant calibration. In Section~\ref{sec:analyses}, we present all our analyses and results, including simulations exploring redundant calibration performance (Section~\ref{sec:omnical_simulation}), inspection of the calibration solutions themselves (Section~\ref{sec:solutions}) and PS analyses (Section~\ref{sec:PS}). We further discuss over these analyses and present our conclusions in Section~\ref{discussion}.

\section{Observations}
\label{sec:obs}

The Murchison Widefield Array (MWA) is a low-frequency (80-300 MHz) radio interferometer located at the Murchison Radio-astronomy Observatory (MRO) in Western Australia, the location of the future low-frequency Square Kilometre Array (SKA).
In MWA Phase I (2013-2016; \citealt{Tingay_2013,Bowman_2013}), 128 antenna tiles\footnote{Each tile consists of 16 individual dipoles with signals combined in an analog beamformer.} were arranged in a pseudo-random layout designed for excellent instantaneous UV-coverage. The Phase II upgrade \citep{wayth_2018,Beardsley_2019} added 128 additional tiles in mid 2016. The correlator is still limited to processing only 128 tiles at a time, so the array is periodically reconfigured between ``compact'' and ``extended'' configurations. The MWA Phase II compact array consists of 72 tiles in a regular hexagonal grid and 56 pseudo-random tiles in a dense core with minimal redundancy. This configuration allows us to perform redundant calibration while retaining good UV-coverage.
\begin{figure}
\includegraphics[width=84mm]{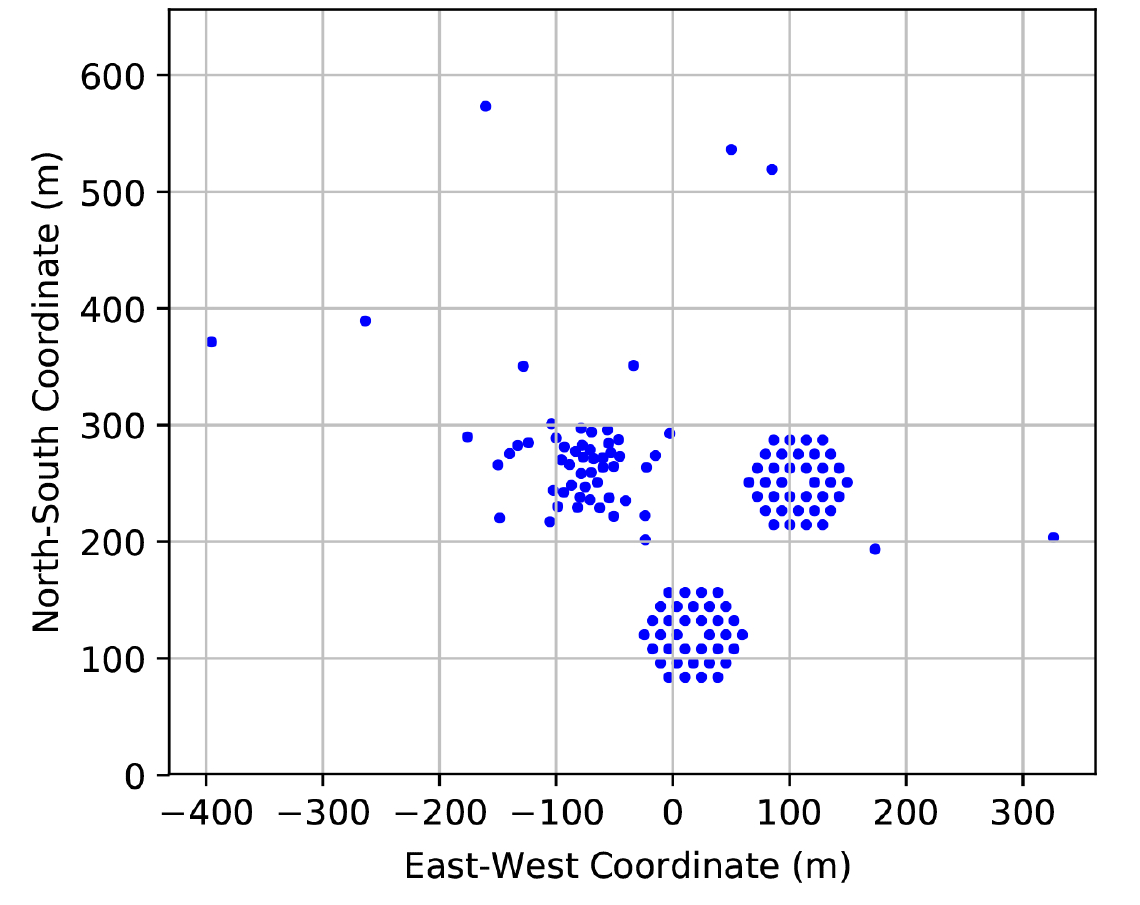}
\caption{The MWA Phase II compact array layout.\label{compact_array}}
\end{figure}

The individual dipoles in each tile can be time-delayed relative to each other in the beamformer, enabling the array to be ``pointed'' to a certain patch of the sky.  However, the time delays can only be adjusted by discrete amounts (as they are added by inserting additional physical lengths of cable into the signal chain), meaning that the telescope cannot continuously track a patch of sky.  Instead, 
the MWA EoR program observes in a specific direction with the sky drifting overhead for about 30 minutes and then re-points the tiles to track the field of interest, which is called the ``drift and shift'' method \citep{Trott_2014}. During an observation, the delays are shifted multiple times until the field is too low in the sky to track, so the data in the observation are grouped into ``pointings'' (each set of tile delays corresponds to one pointing). With the sky transiting from the east, the program begins observing at 5 pointings before zenith and ends observing at 4 pointings after zenith. The zenith is labeled as ``pointing 0'', thus all the 10 pointings are labeled from pointing $-5$ to pointing $4$.

The MWA EoR program focuses on three fields with relatively low foreground emission: 
``EoR0'' ($\text{RA}=0^{\text{h}}00$, $\text{Dec}=-27^{\circ}$), ``EoR1'' ($\text{RA}=4^{\text{h}}00$, $\text{Dec} = -27^{\circ}$), and ``EoR2'' ($\text{RA}=11^{\text{h}}33$, $\text{Dec} = -10^{\circ}$); see \cite{Jacobs_2016} for the motivation for selecting these fields and \cite{trott2020} for a detailed analysis of the features present in each.  Each field fills the main lobe of the MWA primary beam, spanning roughly $\sim 40^{\circ}$ on a side. 

In this work, we analyze high band (167 - 197 MHz) data from three nights observing EoR1 and one night observing EoR0.  The most salient difference between these two fields is the presence of Fornax A, a bright, extended source in the EoR1 field; EoR0, in comparison, contains mostly unresolved sources.  In this work, we denote the nights as ``EoR1 night1'' (2016-10-15), ``EoR1 night2'' (2016-10-17), ``EoR1 night3'' (2016-11-15) and the ``EoR0 night'' (2016-11-21), respectively. 
We explore the four nights using good pointings determined by the window power metric \citep{Beardsley_2016, Wenyang_2019} and the ionospheric metric \citep{Jordan_2017,Trott_2018}:
for EoR1 night1 and night2, good pointings are -3, -2, -1 and 0; for EoR1 night3, good pointings are -3, -2, and -1; for the EoR0 night, good pointings are -2, -1, 0, 1, and 2.



The MWA uses the AOFlagger algorithm\footnote{http://aoflagger.sourceforge.net/doc/api/} to flag data contaminated by RFI \citep{offringa_2015}.
In this work, we further flag faint RFI not captured by the pipeline using data quality metrics developed by \citet{Wilensky_2019} and \citet{Wenyang_2019}. 


\section{Calibration Formalism}
\label{sec:cal}
The sky signal is modified along the path from where it was emitted to where it was recorded. In order to measure the cosmological signal, all these modifications must be corrected for. 
While our RFI flagging and data quality metrics are used to avoid the most corrupted sky signals,
the electronic response of the antennas must still be corrected for with a gain calibration algorithm. 

In this section, we will briefly introduce two calibration methods used in our analysis: sky-based calibration and redundant calibration, and the code packages and algorithms we use to implement them.
\subsection{Measurements of the sky}

Suppose antenna $i$ with gain $g_i$ measures a frequency domain signal $s_i$ at a given instant, an approximation of a short time block. $s_i$ is related to the true sky signal $h_i$ and the antenna's instrumental noise contribution $n_i$ as
\begin{equation}
    s_i=g_i h_i+n_i,
\end{equation}
which is both an equation of time and frequency.

In this paper, we consider only a single linear polarization at a time. We assume per antenna and per frequency gains. In the MWA, the instrumental gains are assumed to be constant over 2-minutes observations. Then in a long integration time that $\langle n_i \rangle=0$, the baseline $b_{ij}$ measures the correlation between signals from antenna $i$ and $j$ ($i\not= j$):
   \begin{subequations}\label{eqn:measure}
   \begin{empheq}{align}
      &v_{ij}=\langle s_i^*s_j\rangle\label{subeqn-1:measure} \\
           &\approx g_i^* g_j \langle h_i^*h_j\rangle+\langle n_i^*n_j\rangle+g_j\langle n_i^*h_j\rangle+g_i^*\langle h_i^*n_j\rangle \label{subeqn-2:measure} \\
           &= g_i^* g_j u_{ij}+n_{ij},\label{subeqn-3:measure}  
   \end{empheq}
   \end{subequations}
where $u_{ij}=\langle h_i^*h_j\rangle$ denotes the true visibility and $n_{ij}$ is the residual noise as the synthesis of all noise contributions in Equation~\ref{subeqn-2:measure}. We have applied two assumptions: 1. the noise of different antennas are uncorrelated; 2. the sky signal doesn't correlate with the noise.  Under these assumptions, the last three terms in Equation \ref{subeqn-2:measure} all have an expectation value of 0 in the limit that time goes to infinity.  However, since our integration length is finite, these three terms contribute an approximately $u_i$-independent Gaussian random component, which we re-label as a single effective noise term $n_{ij}$ in going to Equation \ref{subeqn-3:measure}.

Our goal is to solve for the true sky visibilities $u_{ij}$. Numerous algorithms and strategies have been developed for precise calibration.
Most of them can be classified into two approaches, sky-based calibration and redundant calibration. We will briefly introduce both of the two calibration methods in Section~\ref{CalibrationTech}.

\subsection{Calibration techniques}
\label{CalibrationTech}

In our analysis, we used two calibration packages: one is the Fast Holographic Deconvolution (FHD\footnote{https://github.com/EoRImaging/FHD}; \citealt{barry_2019,sullivan2012fast}) software pipeline, which is used for sky-based calibration, and the other is {\tt\string OMNICAL}\footnote{https://github.com/jeffzhen/omnical} \citep{OMNICAL}, performing redundant calibration.\footnote{Some confusing nomenclature unfortunately exists around this code.  The package \texttt{OMNICAL} contains three algorithms: \texttt{Logcal}, \texttt{Lincal}, and \texttt{Omnical}.  The \texttt{Omnical} algorithm, first described in \cite{dillon2020}, is a more efficient replacement for \texttt{Lincal}; however, following \cite{Li_2018} and \cite{Wenyang_2019}, we use only \texttt{Logcal} and \texttt{Lincal}, which are described below.  In referring to \texttt{OMNICAL} (all capitalized) we refer to the code package and not the \texttt{Omnical} algorithm, which is not used in this work.}

They are two different methods but both work on maximum-likelihood estimate of per-frequency antenna gains. 
For any baseline $b_{ij}$, we assume its noise is Gaussian and mean zero with variance $\sigma_{ij}^2$ \citep{TMS}. 
Maximizing the likelihood function of gains is equivalent to minimizing 
\begin{equation} \label{chisquared}
    \chi^2=\sum_{ij}\frac{|v_{ij}-g_i^*g_j u_{ij}|^2}{\sigma^2_{ij}}.
\end{equation}
Instrumental frequency dependence is implicit in this equation.

\subsubsection{FHD: sky-based calibration}
Sky-based calibration works by substituting model visibilities for the true visibilities in Equation~\ref{chisquared}. The accuracy of the model visibilities determines the accuracy of sky-based calibration. We use GLEAM \citep{GLEAM}, an extragalactic point source catalogue with large coverage and high completeness, as our model sky to generate model visibilities. FHD simulates all reliable sources in the catalogue out to typically 1\% beam level in the primary lobe and the sidelobes to build a theoretical sky \citep{barry_2019,Beardsley_2016}. In this work, we exclude baselines shorter than 50 wavelengths (which are sensitive to diffuse Galactic emission absent from GLEAM) to reduce the contamination from differences between the model visibilities and the true visibilities \citep{barry_2019}.

FHD calibration begins with a set of estimated initial solutions for ${g_i^*(f)}$ 
and the sky visibilities are substituted with model visibilities $m_{ij}$. (In practice, the digital gains in the MWA correlator pre-scale the visibilities to roughly the right level --- so an initial gain estimate of 1.0 at for all antennas at all frequencies is sufficient for quick convergence of the calibration algorithm.)  Then minimizing Equation~\ref{chisquared} becomes a linear least-squares problem:
\begin{equation}
    \chi_{j,sky}^2(f)=\sum_{i}\frac{|v_{ij}(f)-g_i^*(f)g_j(f) m_{ij}(f)|^2}{\sigma^2_{ij}(f)},\label{skychi}
\end{equation}
where $m_{ij}$ is the model visibility. 
This relation can be solved iteratively to return the values of the $g$ terms that minimize $\chi^2$ \citep{Mitchell_et_al,Salvini2014StEFCalA}.


Sky-based calibration requires an accurate sky model, otherwise the difference between the true sky visibilities and model visibilities will be incorporated into our evaluations of instrumental gains. \cite{Barry_2016} demonstrate how gain errors due to sky model incompleteness introduce contamination that prevents recovery of an EoR signal.

\subsubsection{{\tt\string OMNICAL}: redundant calibration}

If an array is built to have sufficient redundant baselines, the total number of measurements will exceed the sampled points in Fourier space. In this case, our equations are over determined and we are able to directly solve for both instrumental gains and visibilities \citep{Liu_2010}. Pedagogically, we can understand this calibration method as minimizing 
\begin{equation}\label{redchi}
    \chi_{\rm red}^2(f)=\sum_{\alpha}\sum_{\{i,j\}_{\alpha}}\frac{|v_{ij}(f)-g_i^*(f)g_j(f) u_{\alpha}(f)|^2}{\sigma^2_{ij}(f)},
\end{equation}
where $\alpha$ indexes types of redundant baselines and $\{i,j\}_{\alpha}$ are sets of antennas that belong to each type of redundant baseline.  $\{u_{\alpha}\}$ and $\{g_i\}$ can both be solved for directly.

In this work, redundant calibration incorporates two parts, {\tt\string Logcal} and {\tt\string Lincal}. {\tt\string Logcal} uses logarithms to cast the minimization of Equation~\ref{redchi} as a linear regression problem. But this method is biased for two reasons: 1. when the signal noise ratio (SNR) is low, it's less accurate to expand the logarithm noise contribution term $\ln \left(1+n_{ij}/g_i^*g_ju_{ij}\right)$ as $n_{ij}/g_i^*g_ju_{ij}$, as is done by {\tt\string Logcal}; 2. {\tt\string Logcal} applies least squares estimation to the distributions of noises' magnitudes and phases, which is inaccurate as the magnitudes and phases of noise follow Rayleigh distributions rather than Gaussian distributions.

{\tt\string Lincal} was developed to mitigate these biases \citep{Liu_2010}. In {\tt\string Lincal}, we Taylor expand Equation~\ref{subeqn-3:measure}: 
\begin{equation}\label{lincal}
v_{ij}\approx g_i^{0*}g_j^0u_{ij}^0+g_i^{0*}u_{ij}^0\Delta g_j+g_j^{0}u_{ij}^0\Delta g_i^*+g_i^{0*}g_j^0\Delta u_{ij},
\end{equation} 
where $\Delta g_j$, $\Delta g_i^*$, and $\Delta u_{ij}$ are equal to $g_j-g_j^0$, $g_i^*-g_i^{0*}$ and $u_{ij}-u_{ij}^0$, respectively. Thus the measurement equation becomes linear so that we can perform a least-squares fitting: using initial guesses of $g_i^0$ and $u_{ij}^0$, we can solve for all values of $(g_i-g_i^0)$ and $(u_{ij}-u_{ij}^0)$. Then we use the new estimations for $g_i^0$ and $u_{ij}^0$ as inputs, and solve the equation iteratively.  Note that the initial inputs should be in the minimum containing the true solution. In this work, outputs of {\tt\string Logcal} are used as the starting point for {\tt\string Lincal}, so that biases are expected to be mitigated.
 
 Redundant calibration faces another problem, in that there are four intrinsic degeneracy parameters per frequency: the overall amplitude, overall phase, and two phase gradient components \citep{Byrne_2019,Dillon_2018,Liu_2010}. These parameters can take any value without affecting the redundancy of the data; therefore, they cannot be constrained by redundancy-based methods. 
We constrain these degenerate parameters using the FHD calibration solutions: finding values of the 4 degeneracy parameters per frequency per polarization for the whole array, which best fit the sky-based calibration solutions \citep{Li_2018}. \cite{Li_2018} explored different ways of combining FHD and {\tt\string OMNICAL}, each of which used slightly different methods for constraining the degeneracy parameters.
 We use the method referred to as ``FOCal'' (also used in \citealt{Wenyang_2019}), which we now detail in Section~\ref{hybrid_cal}.

\subsubsection{FHD and {\tt\string OMNICAL} in tandem: redundant plus sky-based calibration }
\label{hybrid_cal}

In this paper, we perform redundant plus sky-based calibration using FHD and {\tt\string OMNICAL} in tandem. Following \cite{Li_2018}, we perform FHD sky-based calibration on the raw data first. Then we perform redundant calibration on FHD calibrated data, yielding a second set of gains that can be multiplied by the FHD-produced gains to yield a final set of calibration solutions.  Before multiplying the {\tt \string OMNICAL} and FHD gains together, however, we set the parameters in the redundant calibration degenerate space of the {\tt \string OMNICAL} gains to 1 (for the amplitude) and 0 (for the phase terms) so that the FHD values in this subspace are retained after multiplication.  This is the process referred to as ``degeneracy projection'' in \cite{Li_2018}. 
With this approach, we can investigate the performance of each calibration method by looking at both the solutions produced by FHD and the changes to those solutions produced by {\tt \string OMNICAL}.



One can expect in our calibration pipeline that based on FHD calibrations, redundant calibration makes further modifications to fit the instrumental redundancy. These modifications on top of FHD calibration results will be small if sky-based calibration works well; based on this expectation, we refer to the gain solutions produced by {\tt \string OMNICAL} as multiplicative ``$\Delta$'s'' on top of the FHD solutions (i.e. $g_{\rm tandem} =\Delta_{\tt OMNICAL} \times g_{\rm FHD}$). 
In Section~\ref{sec:analyses}, we will look at FHD gain solutions, the redundant calibration gain $\Delta$'s and the difference in the PS when using only FHD vs FHD plus {\tt\string OMNICAL} tandem calibration.






\section{Analyses}
\label{sec:analyses}

Our approach for using FHD and {\tt\string OMNICAL} in tandem enables us to study the performance of each algorithm relatively straightforwardly. Assuming perfect sky-based calibration with a perfect sky model, the $\Delta's$ produced by redundant calibration should be 1.0 (since the gains are multiplicative). Any deviation from 1 indicates an error---either one in sky-based calibration that is being corrected by redundant calibration, or one in redundant calibration due to e.g. non-redundancy in baselines that were assumed to be redundant. Theoretically, the errors in our sky models should vary from field-to-field on the sky, so that sky-based calibration will behave differently for different observations. 
However, it is hard to unambiguously evaluate ``effectiveness'' in real data, when non-unity values for the redundant calibration $\Delta$'s can result from errors in either calibration algorithm.  In this work, we calibrate and process MWA Phase II data from both the EoR1 field (which has not been studied in this fashion before) and the EoR0 field (which was studied in \citealt{Li_2018} and \citealt{Wenyang_2019}). By comparing both the calibration solutions from each algorithm and the resultant PS measurements, we can begin to understand how the behavior of these algorithms varies from field-to-field.

In this section, we will present three analyses: a simulation exploring redundant calibration performance on noisy but otherwise perfect data (\S\ref{sec:omnical_simulation}); a study of the calibration solutions from both FHD and \texttt{OMNICAL} and their repeatability from night-to-night (\S\ref{sec:solutions}); and an investigation of the effects applying the redundant calibration solutions have on the power spectrum of our data (\S\ref{sec:PS}).

\subsection{Simulation: redundant calibration performance on noise}
\label{sec:omnical_simulation}

\begin{figure*}
\includegraphics[scale=0.58]{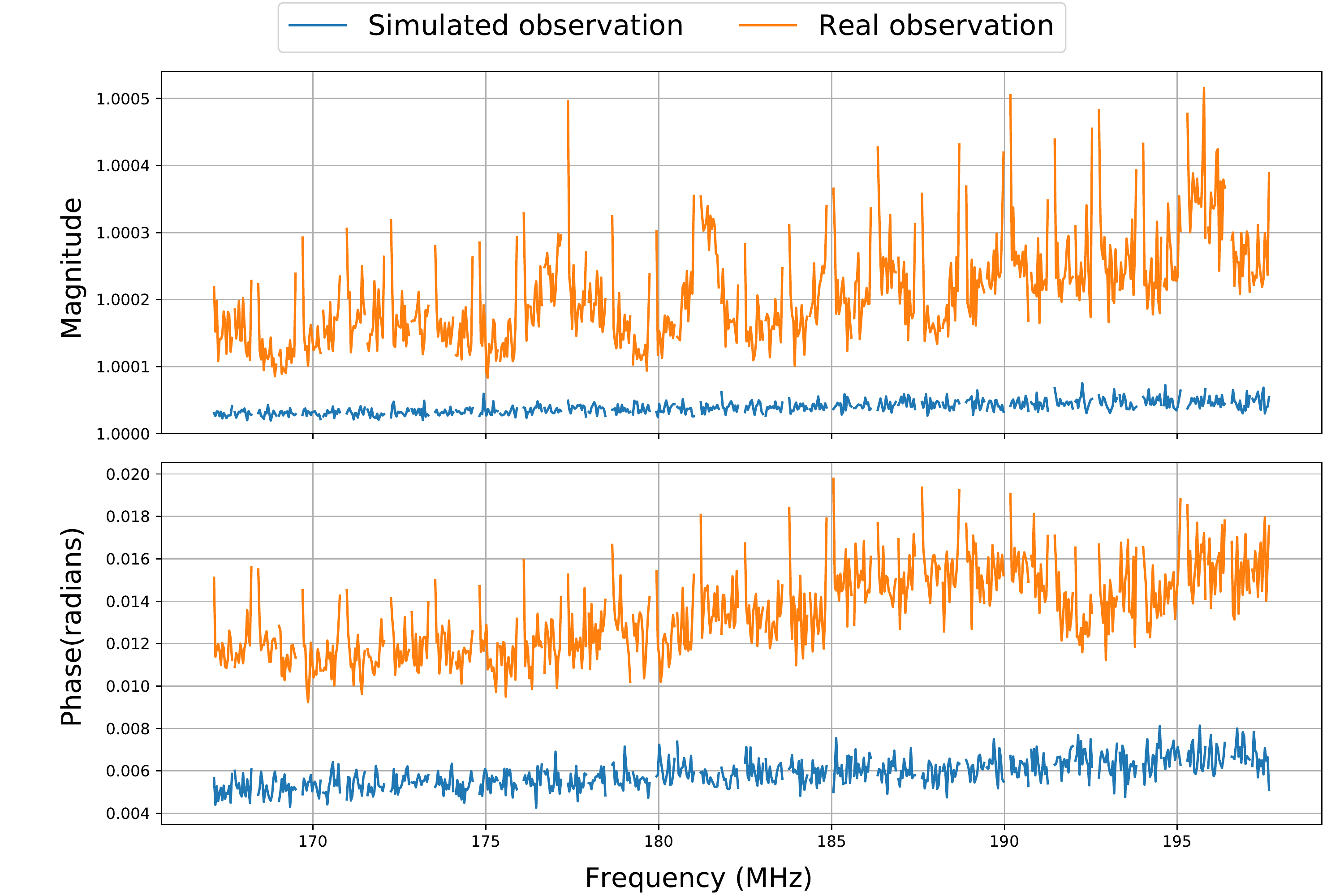}
\caption{The RMS of redundant calibration solution magnitudes/phases of a real/simulated observation. Orange lines represent the real EoR1 observation 
and blue lines correspond to the simulated observation, which is constructed to have the same mean noise as the real observation per frequency and the same total SNR averaged over frequencies. In addition, we set the SNR of the simulated observation proportional to frequency to -1.59 which is similar to the real one. 
Upper: RMS of redundant calibration solution magnitudes; Bottom: RMS of redundant calibration solution phases.
We can see that redundant calibration makes changes to the real data beyond just noise.
\label{real_fake}}
\end{figure*}


Since our measured data has a random noise component, we can never truly expect redundant calibration to return $\Delta$'s of exactly 1.0, even if sky-based calibration is perfect.  
Here, we present a simple simulation where {\tt\string OMNICAL} is run on ``perfect'' data with a signal-to-noise ratio (SNR) comparable to the real observations.  If we find the redundant calibration $\Delta$'s for the real data are at a level of order the $\Delta$'s we find in this simulation, we might conclude that redundant calibration is purely fitting noise, i.e., any changes it wants to make to the FHD calibration are purely random.

To keep the simulation simple, we create a single mock observation with perfect redundancy and no fringes, i.e., the signal is real valued. For the fairness of comparison, the simulated observation is designed to have the same noise level and comparable SNRs with the real observation at each frequency. We characterize the noise level via the standard deviation of the noise and SNR is defined as the ratio between signal amplitude and the noise level. The frequency dependent noise level has been assumed to be constant over all baselines and over an observation. Thus we can produce a simulated observation such that the signal $\mu(f)$ and the standard deviation of the noise $\sigma(f)$ are the same over all baselines and times per frequency.
Mathematically, the simulated visibility $V$ is
\begin{equation}
\label{eq:simvis}
    V=\mu+n=\text{SNR}(f)\cdot \sigma(f) + \text{N}_R\left(0,\sigma^2(f)\right) + j\text{N}_I\left(0,\sigma^2(f)\right),
\end{equation}
where $n$ denotes the complex noise contribution of the simulated visibility, with standard deviation $\sigma(f)$ in the real and imaginary, and $\mu=\text{SNR}(f)\cdot\sigma(f)$. Both the real and imaginary noise components are randomly sampled as mean zero Gaussian $\text{N}\left(0,\sigma^2(f)\right)$. 

We directly calculate the per frequency noise levels of the real observation and use them as $\sigma(f)$ in the simulated observation. 
The $\text{SNR}(f)$ is also empirically calculated from the real observation. 
The noise level is calculated per-frequency for the entire observation using the even/odd differencing method described in \cite{barry_2019}: visibilities separated by two seconds in time are subtracted from each other to remove the slowly-changing sky signal.  The residuals are then assumed to be noise dominated, with a per-frequency standard deviation (taken across time and baseline) that reflects the underlying noise in the data.



The ``signal'' level of the real data is calculated from the distribution of the visibility amplitudes for a single observation.
For a set of visibilities with constant sky signal magnitude $\mu$ and noise level $\sigma$, the distribution of visibility amplitudes $Z$ is uniquely determined, even though the visibilities themselves rotate through the complex plane as a function of time. 
The analytic expression for the mean value of visibility amplitudes can be shown to be:
\begin{equation}\label{mapping_amplitudes}
\begin{split}
\left\langle Z\right\rangle&=\int_0^{\infty}Z\cdot P(Z)dZ=\frac{e^{-\frac{\mu^2}{2\sigma^2}}}{\sigma^2}\int_0^{\infty}Z^2 e^{-\frac{Z^2}{2\sigma^2}}\textbf{I}[0,\frac{Z\mu}{\sigma^2}]dZ\\
&=\sqrt{\frac{\pi}{8}}\sigma^{-1} e^{-\frac{\mu^2}{4\sigma^2}}\left((2\sigma^2+\mu^2)\textbf{I}[0,\frac{\mu^2}{4\sigma^2}]+\mu^2\textbf{I}[1,\frac{\mu^2}{4\sigma^2}]\right),\\
\end{split}
\end{equation}
where $\textbf{I}[\nu, z]$ denotes modified Bessel function of the first kind:
\begin{equation}
\textbf{I}[\nu, z]=\sum_{k=0}^{\infty}\frac{1}{\Gamma(k+\nu+1)k!}\left(\frac{1}{2}\right)^{2k+\nu}.
\end{equation}
This relation lets us calculate a value of $\mu$ to use in our simple visibility model (Equation \ref{eq:simvis}) and $\left\langle Z\right\rangle$ of the real observation. For the real observation we used in comparison, $\left\langle Z\right\rangle \approx 2.707\sigma$, giving us $\mu\approx 2.495\sigma$.

We calculated the mean SNR of the real observation over all frequencies and then produce a series \{SNR($f$)\} whose mean value is equal to the real observation's such that SNR($f$) is proportional to frequency to the -1.59, which is derived from fitting a power-law distribution to the empirically measured frequency dependence of SNR.
The simulated observation with known SNR($f$) is directly input into {\tt\string OMNICAL} so that the $\Delta$ values should be mean 1 with any deviations purely caused by noise.

We compare $\Delta$'s of a real EoR1 observation (Observation ID: 1163254256) with redundant calibration solutions of the corresponding simulated observation in Figure \ref{real_fake}, which shows root mean square (RMS) of redundant calibration solution magnitudes/phases. The RMS is taken across the set of antennas. Clearly, redundant calibration is making changes to the real data at a level significantly larger than might be expected were the data perfectly redundant up to the noise level.

We also investigate the effect of SNR on the result of the simulations.  We find that as long as the SNR is held fixed (e.g. increases to the signal strength are offset by increases to the noise), the $\Delta$'s produced by redundant calibration are at the same level.  Increasing the SNR does reduce the magnitude of the $\Delta$'s, indicating that redundant calibration is indeed affected by noisy data.  However, as indicated in Figure \ref{real_fake}, the RMS of the $\Delta$'s for the real data are larger than would be expected from SNR alone.


From these results, we can conclude that {\tt \string OMNICAL} is responding to real non-redundancy in the data that is above the noise level.  Our simulation tells us the expected scale of the $\Delta$'s for a given SNR and perfect redundancy; in the real data, we see $\Delta$'s a factor of $\sim 3-5$ larger in RMS.  This analysis does not, however, tell us the source of the non-redundancy.  It might be that the FHD solutions are not completely correct and that {\tt \string OMNICAL} is bringing the gains closer to their true values.  However, it might be that the instrument itself is non-redundant and {\tt \string OMNICAL} is introducing further errors into the gains (since it assumes perfect redundancy).  We now continue our analysis of the gain solutions produced by FHD and {\tt \string OMNICAL} to see whether the addition of {\tt \string OMNICAL} does in fact lead to improvements in our calibration and subsequent analysis.

\subsection{Comparison of calibration solutions\label{sec:solutions}}

To study the performance of FHD and {\tt\string OMNICAL} when applied to actual data, we compare the calibration solutions among the four nights.  We perform two distinct analyses: a detailed investigation of the spectral behavior of four sets (one per pointing that passes the cuts described in \S\ref{sec:obs}) of calibration solutions from 2-minutes LST-matched data sets (\S\ref{sec:lst_matched_obs}) and a study of the repeatability of the calibration solutions across all observations in our data set (\S\ref{sec:all_obs}).

\subsubsection{Spectral Behavior of the Calibration Solutions}
\label{sec:lst_matched_obs}

\begin{figure*}
\includegraphics[width=175mm]{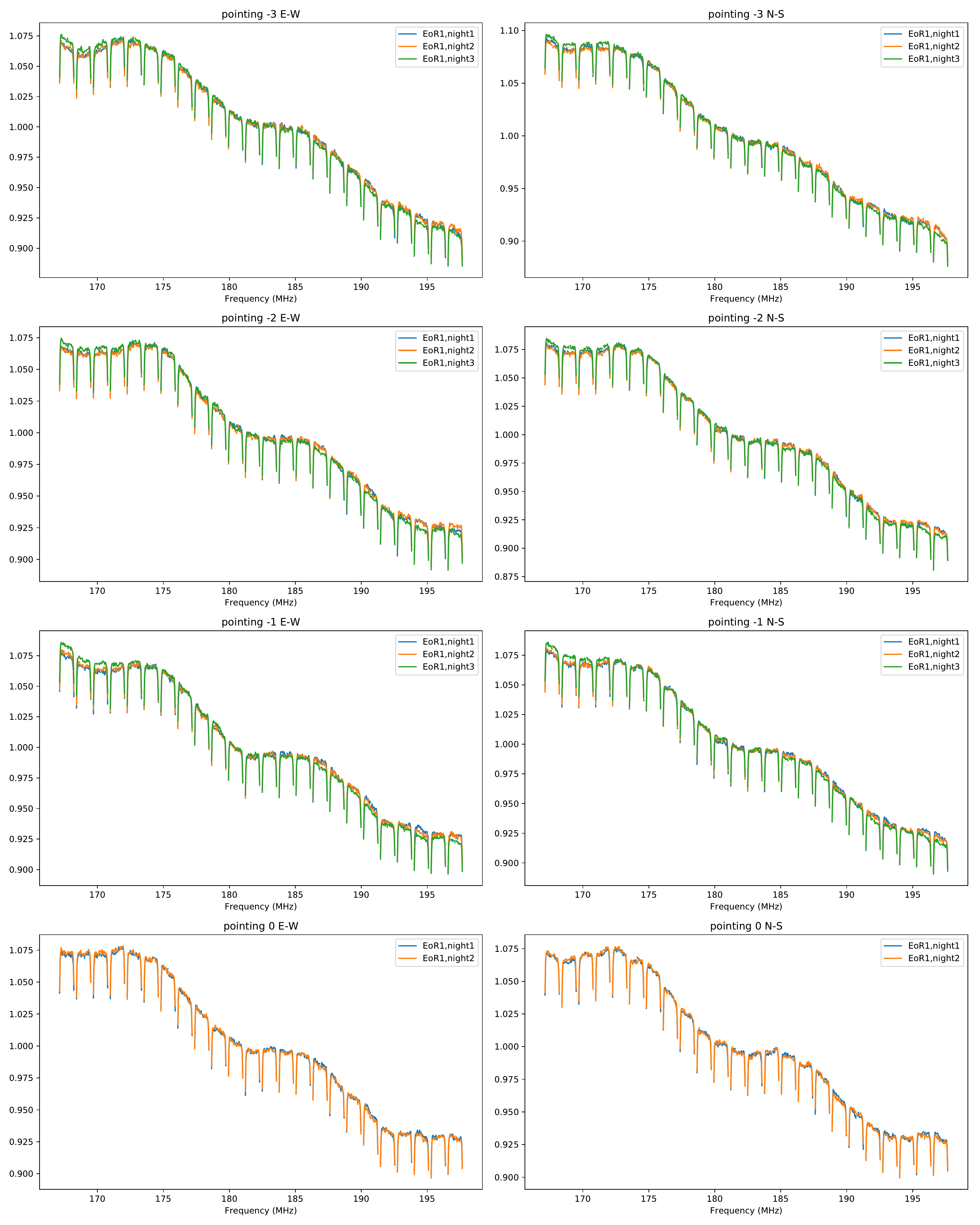}
\caption{The spectra of FHD solution magnitudes (averaged over antennas and normalized to mean one) of observations with well matched LSTs over our three EoR1 nights. Blue, orange and green represent EoR1 night1, night2 and night3, respectively. Two columns correspond to two polarizations and four rows correspond to different pointings. Note that EoR1 night3 pointing 0 is flagged due to ionospheric contamination.  The periodic 1.28\, MHz scalloping is due to the two-stage Fourier transform used by the MWA correlator. 
For each 2-minutes LST-matched data set, the solutions are largely in agreement with only small discrepancies over nights. In general, EoR1 nights 1 and 2 agree closely, with night 3 showing a steeper spectral slope.
\label{fhd}}
\end{figure*}

\begin{figure*}
\includegraphics[width=175mm]{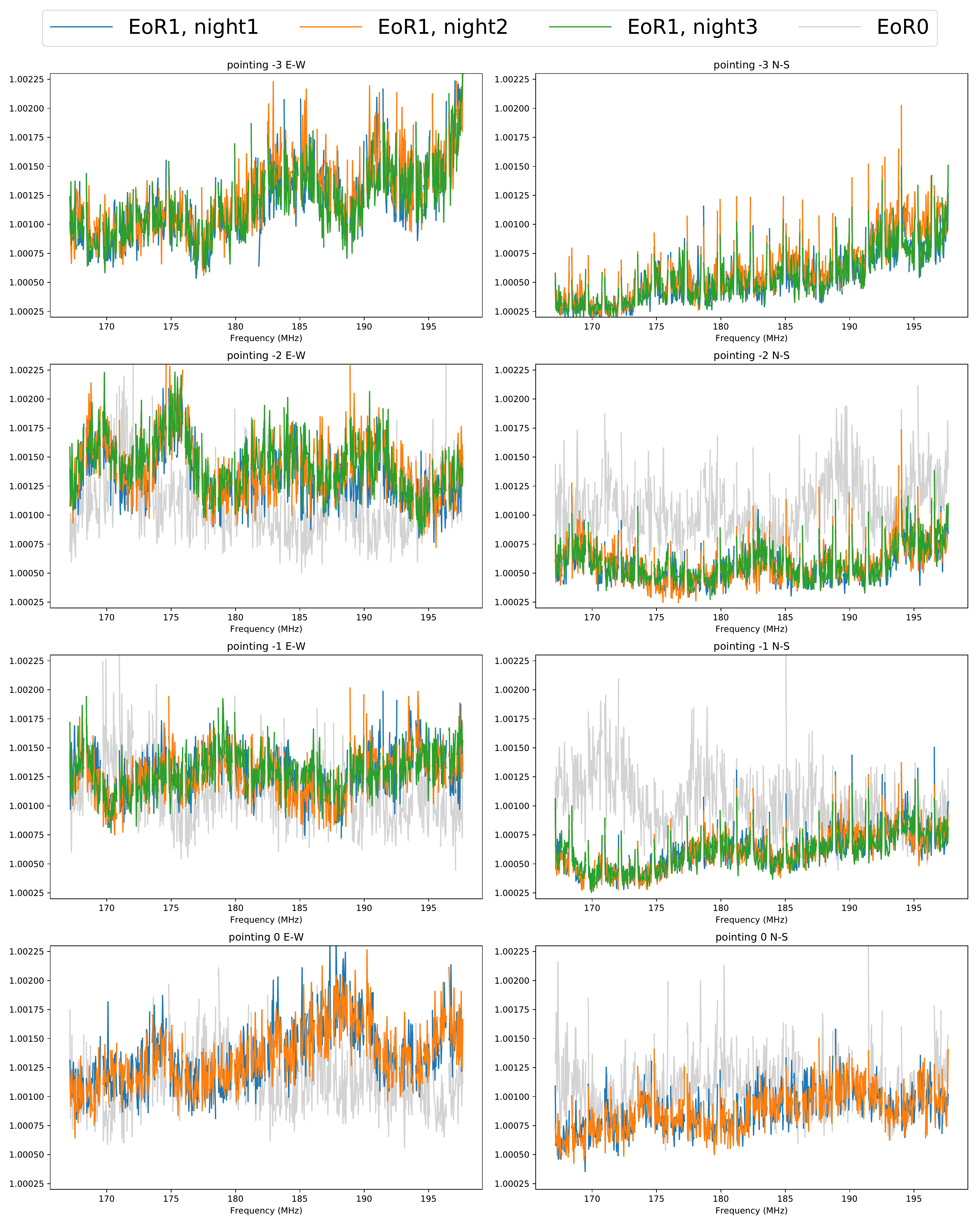}
\caption{The RMS of magnitudes of $\Delta$'s for observations sampled per pointing per polarization and per night. EoR1 observations are sampled to have well-matched LSTs over EoR1 nights. We also sampled typical EoR0 observations at pointing -2, -1, and 0. Four different colors denote different nights (see legend at the top of the figure for details). The columns represent the two linear polarizations and each of the four rows correspond to different pointings. We find that $|\Delta|$'s are consistent at any pointing over all the EoR1 nights with repeatable frequency structure, and the results from EoR0 observations have different spectral behavior. It also shows $|\Delta|$'s for EoR1 observations have bigger values at E-W than at N-S (see further discussion in Section~\ref{discussion}). 
\label{omni_solutions_EoR1_nights}}
\end{figure*}
Within each pointing, we select one observation per EoR1 night such that the sampled EoR1 observations are at well matched LSTs. Similar LSTs will result in similar model visibilities generated by FHD. Thus, comparing LST-matched EoR1 observations across nights will test the consistency and repeatability of the calibration techniques. 

We also select one observation per pointing from the EoR0 night and compare $\Delta's$ with sampled EoR1 observations to explore redundant calibration performance at different EoR fields and a possible sky model dependence. There is obviously no LST accordance between the EoR0 night and EoR1 nights, but the observations chosen are representative.

We begin by first looking at the FHD sky-model calibration solutions for each of our 3 selected EoR1 observations.  We normalize the magnitudes of per frequency FHD solutions for sampled EoR1 observations to mean 1 and then average over antennas to study the typical spectral structure.
Figure~\ref{fhd} shows the magnitude of FHD solutions for sampled observations on the EoR1 nights. For each set of observations with similar LSTs, the solutions are largely in agreement with only small discrepancies over nights. In general, EoR1 nights 1 and 2 agree closely, with night 3 showing a steeper spectral slope.

Comparison of redundant calibration solutions are carried out in a similar way. We look at the spectra of the RMS of the magnitudes of the $\Delta$'s over antennas for each sampled observation. Figure~\ref{omni_solutions_EoR1_nights} illustrates that $|\Delta|$'s are similarly consistent over all the EoR1 nights with repeatable frequency structure.  The gray curve shows the results from an EoR0 observation, which has different spectral behavior, but no significant difference in overall magnitudes (recall that -3 is a flagged pointing for EoR0 due to Galactic contamination). 
A comparison of the East-West (E-W) polarization and the North-South (N-S) polarization shows that $|\Delta|$'s for EoR1 observations have bigger values at E-W than at N-S. We will return to this point in Section~\ref{discussion}, but note that the polarization dependence of redundant calibration solutions is significantly weaker in the EoR0 observations. 

This result suggests that the spectral structure of our calibration solutions are quite repeatable for the same LST on different nights, but there is a moderate dependence on the pointing.  While not shown, we have repeated this analysis for a number of other 2-minutes LST-matched data and we find the results to be very similar.  For redundant calibration solutions, we find a significant difference between the EoR1 and EoR0 fields. 

\subsubsection{Repeatability of Redundant Calibration Solutions}
\label{sec:all_obs}

\begin{figure*}
\includegraphics[width=175mm]{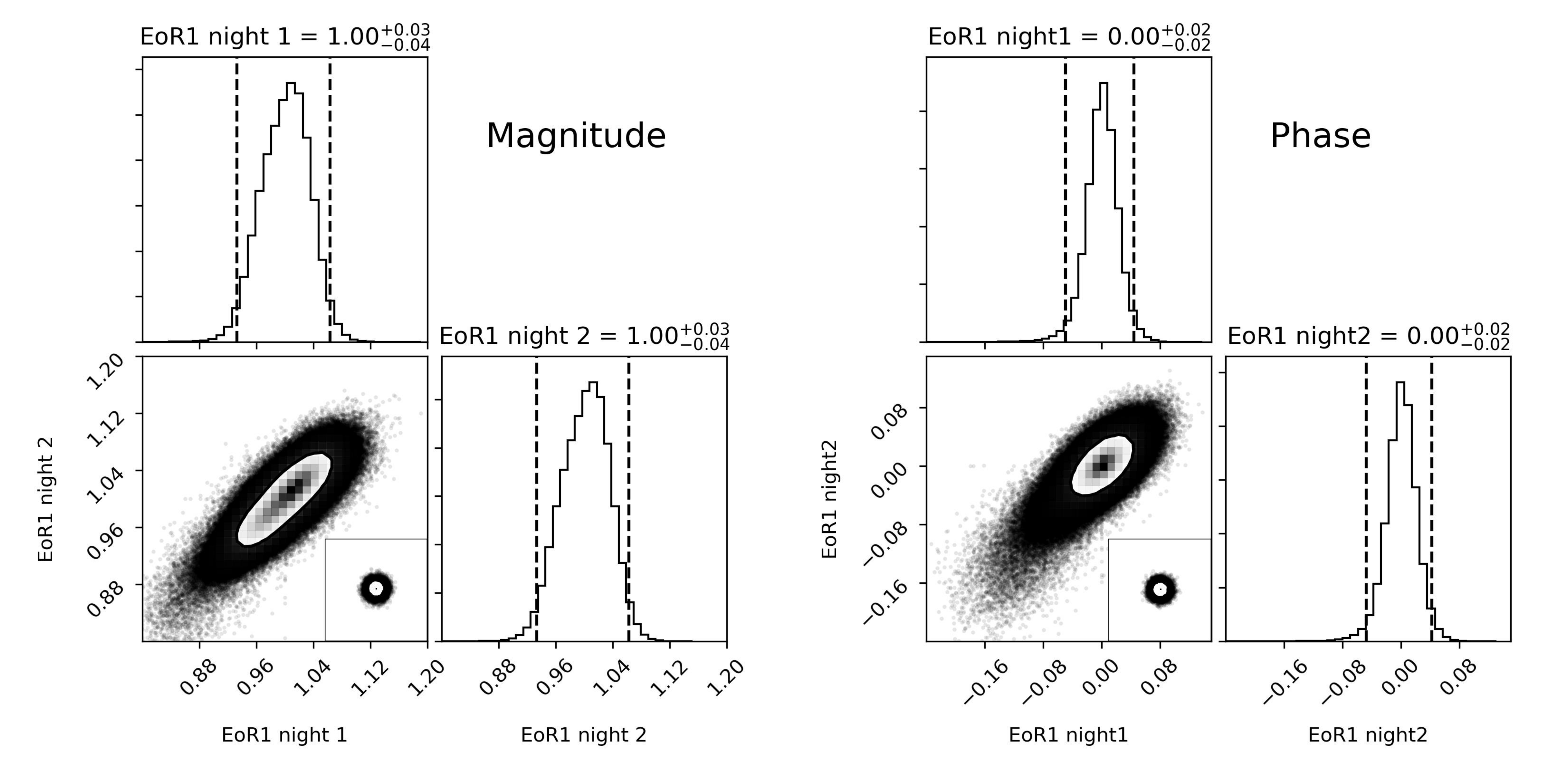}
\caption{Histograms of redundant calibration solutions for EoR1 night 1 and night 2. 
Left: 1 and 2D histograms of the magnitudes of redundant calibration solutions for EoR1 night 1 and night 2.  The subplot in the lower left panel is the 2D histogram of the magnitudes of redundant calibration solutions for two \emph{simulated} data sets with common SNR.
Right: 1 and 2D histograms of the phases (in radians) of redundant calibration solutions for EoR1 night 1 and night 2.
The subplot in the lower left panel is the 2D histogram of the phases of redundant calibration solutions for two \emph{simulated} data sets with common SNR.
For 1D histograms, the area between the vertical dashed lines contains 95\% of the samples; for 2D histograms, a contour (where the plot transitions from a point-by-point scatter plot to a pixelated density plot) contains $1-e^{-2}\approx 84\%$ of the samples.
Comparison with the simulated data shows that there is a scatter larger than would be expected from noise, EoR1 night 1 and night 2 are overall significantly correlated. 
 \label{calibration_repeatability_correlation}}
\end{figure*}

To make a more robust claim about the night-to-night repeatability of redundant calibration, we pair all the LST-matched redundant calibration $\Delta$'s between two EoR1 nights (up to $10^{10}$ pairs in total, with one pair for each frequency, antenna, and time sample of the data set) and study the correlation between the two nights.
Figure~\ref{calibration_repeatability_correlation} shows the results of this study.  The top row of two three-panel ``triangle" plots show the 1- and 2D histograms of all the redundant calibration $\Delta$'s in magnitude (left) and phase (right).  Within each triangle, the 1D histograms show the $\Delta$'s from night 1 (top left) and night 2 (bottom right); the 2D contour/scatter plot (bottom left) shows the values from night 1 plotted against the values from night 2.  
For 1D histograms, the area between the vertical dashed lines contains 95\% of the samples; for 2D histograms, a contour (where the plot transitions from a point-by-point scatter plot to a pixelated density plot) contains $1-e^{-2}\approx 84\%$ of the samples.

The elliptical shape of the 2D histograms, centered on the one-to-one line, suggest that the solutions found on EoR1 night 1 and EoR1 night 2 are highly correlated.  To better assess the significance of this correlation, we perform a similar analysis using two sets of simulated data with common SNR.  Since the simulated data (described in \S\ref{sec:omnical_simulation}) are effectively perfect up to the noise, we expect no correlation in the $\Delta$'s between the two nights (i.e., any change redundant calibration wants to make to the data set is merely a result of fitting noise).  The results of this simulation are shown in the bottom two triangle plots in Figure~\ref{calibration_repeatability_correlation}.
As expected, the 2D histograms are completely circular --- meaning there is no correlation in the $\Delta$'s between the two simulated nights.  The width of these circles also gives us an estimate of the scatter we might expect in perfectly correlated data (because the simulations have an SNR matched to the data).  Comparison between real and simulated data shows that the two EoR1 nights are not perfectly correlated up to the noise; rather, the scatter in these plots (effectively, the minor axis of the ellipses) is larger than would be expected from noise alone.   However, the overall correlation is quite significant: the redundant calibration solutions for the two nights of data have Pearson correlation coefficients of 0.826 (magnitude) and 0.651 (phase),  confirming our intuition that redundant calibration is producing largely repeatable results from night-to-night.


\subsection{Comparison of Power Spectra}
\label{sec:PS}

\begin{figure}
    \includegraphics[width=\columnwidth, trim=0 0.7cm 0 0.7cm]{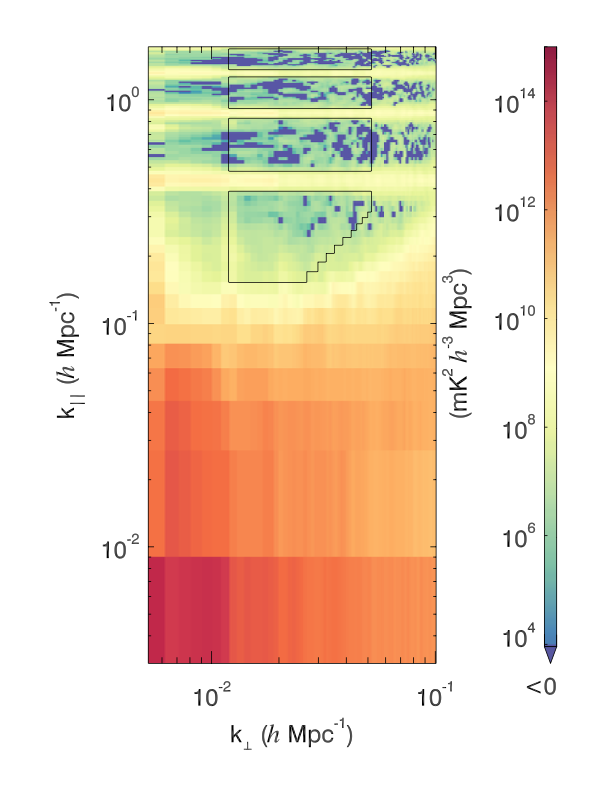}
    \caption{An example for 2D PS plot to highlight modes that will be used for 1D power in $k$ in Figure \ref{ps1}, \ref{ps0}, and \ref{ps2}. 
    To avoid coarse band contamination, we discard 5 $k_{\parallel}$ bins around the centre of each coarse band mode.
    Low $k_\parallel$ values are inside ``the wedge'' and are therefore not included.
    In the $k_{\perp}$ direction, we keep modes between a lower bound of 12$\lambda$ and an upper bound of 50$\lambda$ \citep{Wenyang_2019}. The cutting is performed in the 3D power spectrum and averaged to one-dimension.}
    \label{fig:k_cut}
\end{figure}

\cite{Li_2018} and \cite{Wenyang_2019} evaluated tandem calibration techniques by quantifying the reduction of power in the EoR window (assumed to be due to residual foreground contamination). We apply that formalism in our analyses. 
To illustrate the difference between the two calibration pipelines, we subtract the PS of data with both FHD and redundant calibration applied from that of the FHD-only applied data in 3D $k$ space.  We then select regions of $k$ space where both foregrounds and the periodic contamination from the 1.28~Mhz band structures are minimized  and bin in spherical annuli to make 1D PS difference plots versus $k$.  We plot the 1D results as $\Delta^2(k) = \frac{k^3}{2\pi^2}P(k)$ to facilitate comparison with the literature.  We perform this exercise for both the N-S polarization and the E-W polarization and for all four nights. We produce PS using the package Error Propagated Power Spectrum with Interleaved Observed Noise ($\varepsilon$ppsilon\footnote{https://github.com/EoRImaging/eppsilon}; \citealt{Jacobs_2016,barry_2019}). 
In 1D PS difference plots, solid lines represent positive values, i.e., the power at that $k$ mode is \emph{reduced} by the application of redundant calibration to FHD calibrated data, while dashed lines represent negative values. In the EoR window, the former are regarded as ``improvements''--- redundant calibration has lowered the contamination in the region expected to be foreground free. 
In Figure~\ref{fig:k_cut}, we illustrate the modes we select in a 2D PS (although recall the differencing and mode selection takes place in 3D $k$ space). For consistency, we use the same $k$ space selections as used in \citet{Wenyang_2019}. The selection gives us a range of $|k|$ modes to look at, although we draw particular attention to the modes in 1D PS difference plots where $k$ is about $0.15 \sim 0.20\ hMpc^{-1}$; these modes have both the largest sensitivity to the EoR signal and come from regions of 2D $k$ space closest to ``the wedge'', where calibration errors can most easily cause contamination \citep{Morales_2019}.

\begin{figure*}
\includegraphics[width=175mm]{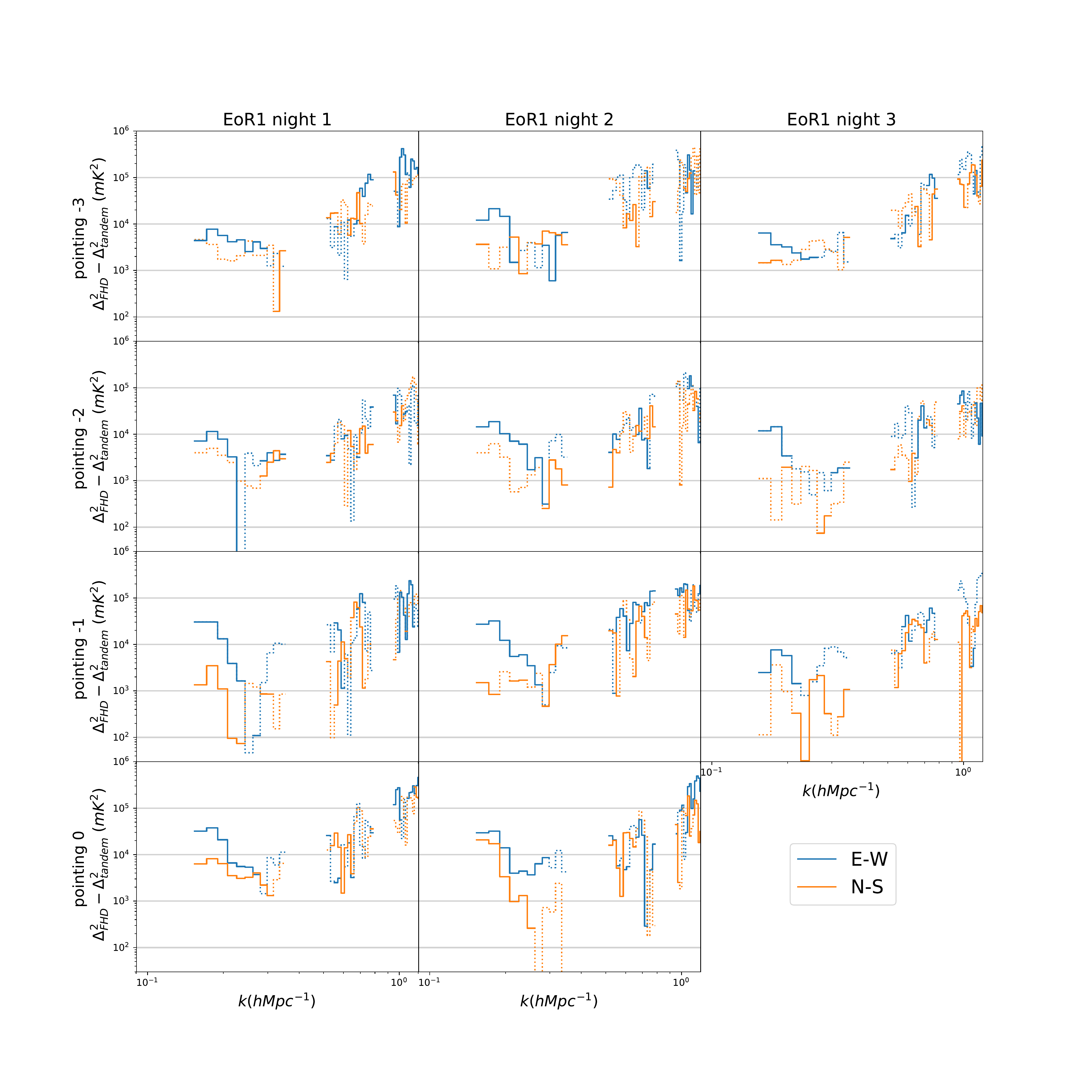}
\caption{ 1D PS difference pointing-by-pointing for the three EoR1 nights. Solid lines represent $k$ modes where redundant calibration has reduced the overall power, while dotted lines imply a negative reduction in power (i.e. redundant calibration has introduced additional contamination). The three columns correspond to the three nights and the four rows correspond to different pointings. Blue represents the East-West polarization while orange is North-South.  Recall that the night 3 zenith pointing is flagged due to ionospheric contamination. For the highest-sensitivity modes ($\sim 0.15 - 0.20\ h{\rm Mpc}^{-1}$), we can see repeatable improvements up to $10^4$\,mK$^2$ for all pointings in the E-W polarization (blue lines), while improvements in the N-S polarization (orange lines) only appear at the zenith pointing (the fourth row).  \label{ps1}}
\end{figure*}

Figure~\ref{ps1} shows the 1D PS difference for each pointing of the EoR1 nights. For each PS plot, we integrate all observations within the pointing not flagged by our quality metrics. For the E-W polarization (blue lines), we can see repeatable improvements up to $10^4$\, mK$^2$ for all pointings.
In contrast, improvements in the N-S polarization (orange lines) only appear at the zenith pointing (the fourth row).
At the off-zenith pointings (first three rows), we see no significant improvements of PS in the N-S polarization.

\begin{figure*}
\includegraphics[width=175mm]{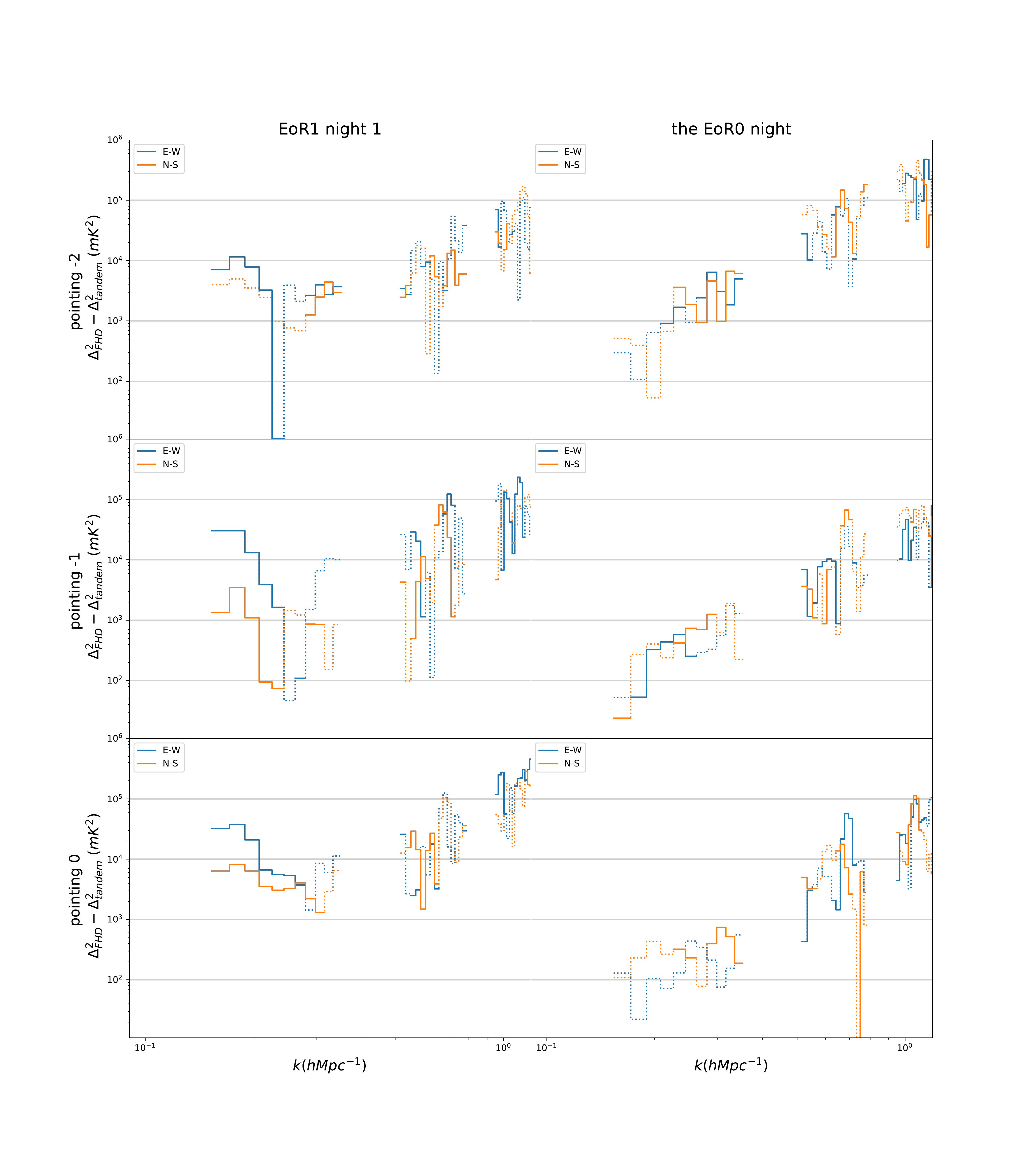}
\caption{ 1D PS difference for EoR1 night 1 (left) and the EoR0 night (right). Solid lines represent modes where redundant calibration has reduced the power, while dotted lines imply an increase. Blue represents the East-West polarization while orange is North-South. The improvements on PS in the East-West polarization are seen for any pointing at all three EoR1 nights, whereas no significant improvements are seen for any pointing of the EoR0 night. No significant improvements in the N-S polarization.\label{ps0}}
\end{figure*}

\begin{figure*}
\centering
\includegraphics[width=175mm]{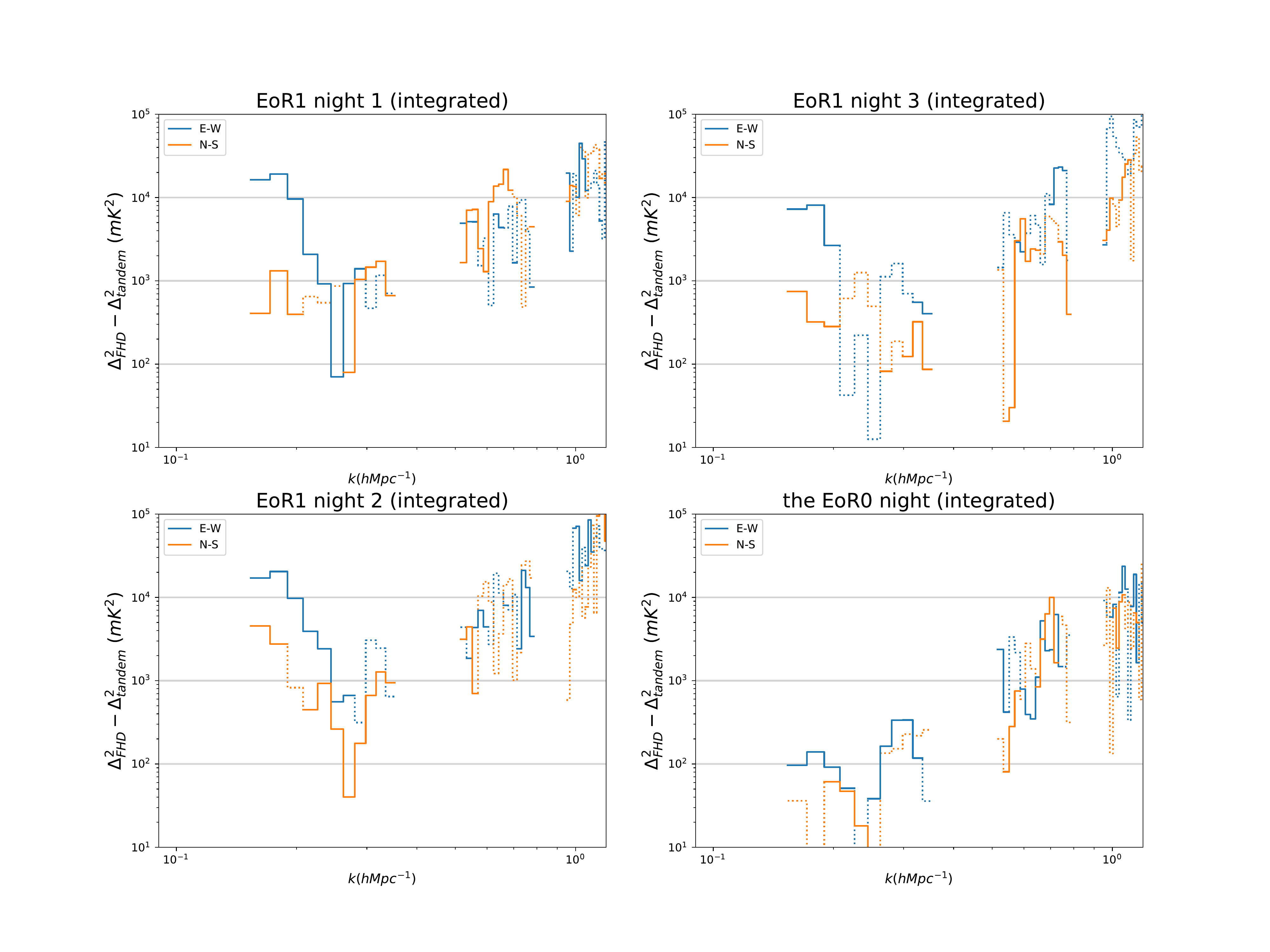}
\caption{1D PS difference of the full night for all nights. Solid lines represent a reduction of power from redundant calibration, while dotted lines imply an increase. Upper left: full night residual PS at EoR1 night1; Bottom left: full night residual PS at EoR1 night2; Upper right: full night residual PS at EoR1 night3; Bottom right: full night residual PS at the EoR0 night. Blue represents polarization East-West while orange represents polarization North-South. The improvements in the integrated, E-W polarization PS are seen for all three EoR1 nights, whereas no significant improvements are seen for EoR0.  Neither field shows significant improvements in the N-S polarization. \label{ps2}}
\end{figure*}

A 1D pointing-by-pointing PS difference comparison of EoR0 and EoR1 is shown in Figure~\ref{ps0}, and Figure~\ref{ps2} shows PS differences after integrating all good pointings for each night. 
 The improvements in the East-West polarization are seen for all three EoR1 nights, whereas no significant improvements are seen for EoR0 in any pointing or in the integrated PS, consistent with the results of \cite{Wenyang_2019}.
 
In general, when we do see consistent improvements, they are at levels of $\sim 10^3$ to $10^4$\,mK$^2$.  Given the recent measurements of \cite{barry_et_al_2019b}, \cite{Wenyang_2019}, and \cite{trott2020}, which use the MWA to achieve lowest limits of $\Delta^2 < 3.9 \times 10^3$\, mK$^2$, $\Delta^2 < 2.39 \times 10^3$\, mK$^2$, and $\Delta^2 < 1.8 \times 10^3$\, mK$^2$, respectively, we can see improvements of this scale can be significant.


\section{Discussion}
\label{discussion}

In Section~\ref{sec:analyses}, we presented analyses looking at both calibration solutions and 1D PS differences. In this Section, we will further discuss two key aspects of our results: 
the comparison of the EoR1 and EoR0 results and the polarization dependence of redundant calibration's improvements in the PS of EoR1 .

\subsection{Comparison of EoR1 and EoR0}
\label{sec:compare_eor0_eor1}

\begin{figure*}
    \centering
    \includegraphics{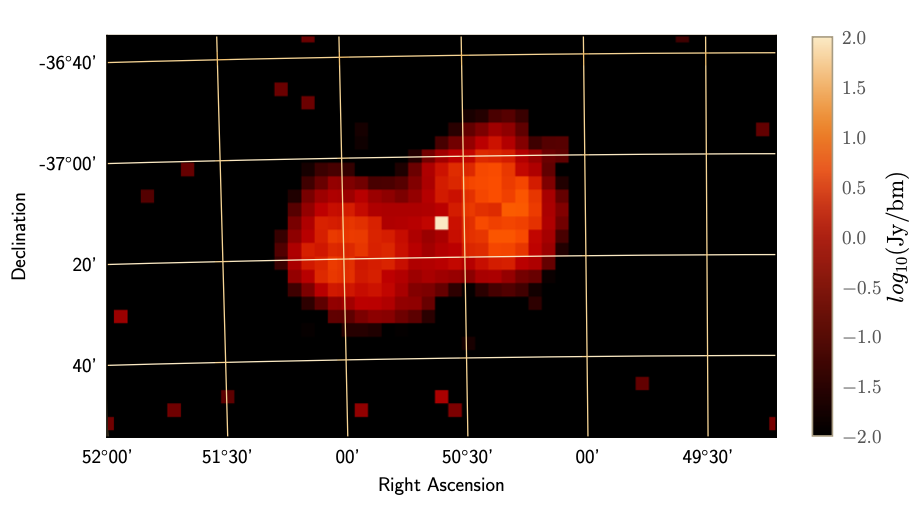}
    \caption{The extended source model used for Fornax A produced using the techniques of \citealt{carroll_et_al_2016}. This image is at 150 MHz and has a peak flux density of $\sim$ 313 Jy. A total of 1925 components are used for this model.}
    \label{fig:Fornax_A}
\end{figure*}

The noise simulations (Section~\ref{sec:omnical_simulation}) and general repeatability of both the gain solutions (Section~\ref{sec:solutions}) and the power spectrum improvements (Section~\ref{sec:PS}) across the three nights of EoR1 give us significant confidence that redundant calibration is indeed constraining real information about the telescope.  It is then interesting to ask why we see no significant PS improvements in our EoR0 analysis (consistent with the results of \citealt{Wenyang_2019}).

The biggest difference between the two fields is the distribution of flux density on the sky.  EoR0 has no significant bright or extended sources \citep{carroll_et_al_2016}, and as such is well-modeled by the GLEAM catalog.  For EoR1, our sky model is likely not as accurate as the EoR0 due to the presence of Fornax A (see Figure~\ref{fig:Fornax_A}) near the center of the field.
GLEAM does not include a Fornax A model \citep{GLEAM} itself; ours is produced directly from Phase I EoR observations using the techniques of \cite{carroll_et_al_2016}.  Figure \ref{fig:polarization} shows the residual images of EoR1 after our sky model has been subtracted; we return to the polarization properties of this image in our discussion in Section~\ref{sec:Polar_diff}, but we generally see a negative feature at the position of Fornax A in the lower right of the image---suggesting our model has an excess of flux density compared to the data.  However, we emphasize that our Fornax A model is by no means ``bad"; the post-subtraction residuals seen in Figure \ref{fig:polarization} are at the percent level compared to the source itself.  \cite{line_et_al_2020} recently produced a new model for Fornax A using shapelets (which, at present, cannot be used by the FHD code), but when subtracted from MWA data, the residuals are at comparable level to those seen here (c.f. their Figure 9).

The fact that Fornax A is so difficult to model leads to interesting conclusions about the role of redundant calibration in the EoR1 field.
If FHD produces calibration solutions with frequency-dependent errors driven by sky-model incompleteness (as suggested in \citealt{Barry_2016}), then we will expect sky-based calibration to perform relatively worse on EoR1; redundant calibration, to the extent that it is sky-model independent, will therefore have a more significant effect. 
Figure \ref{omni_solutions_EoR1_nights} indeed shows that redundant calibration solutions behave differently between EoR1 and EoR0, consistent with the idea that the model errors introduced in FHD are significantly different. 
In fact, even for the observations at the same pointing but separated by several minutes, redundant calibration solutions can have significantly different spectra, which implies that FHD calibration is quite sensitive to LST.
Our results therefore suggest that redundant calibration techniques can still play a role mitigating sky model errors in calibration---especially for fields with difficult, extended sources---even though sky model errors will still enter through the degenerate parameters \citep{Byrne_2019}.



\subsection{Polarization dependence of redundant calibration performance}
\label{sec:Polar_diff}

\begin{figure*}
    \centering
    \includegraphics[scale=0.5]{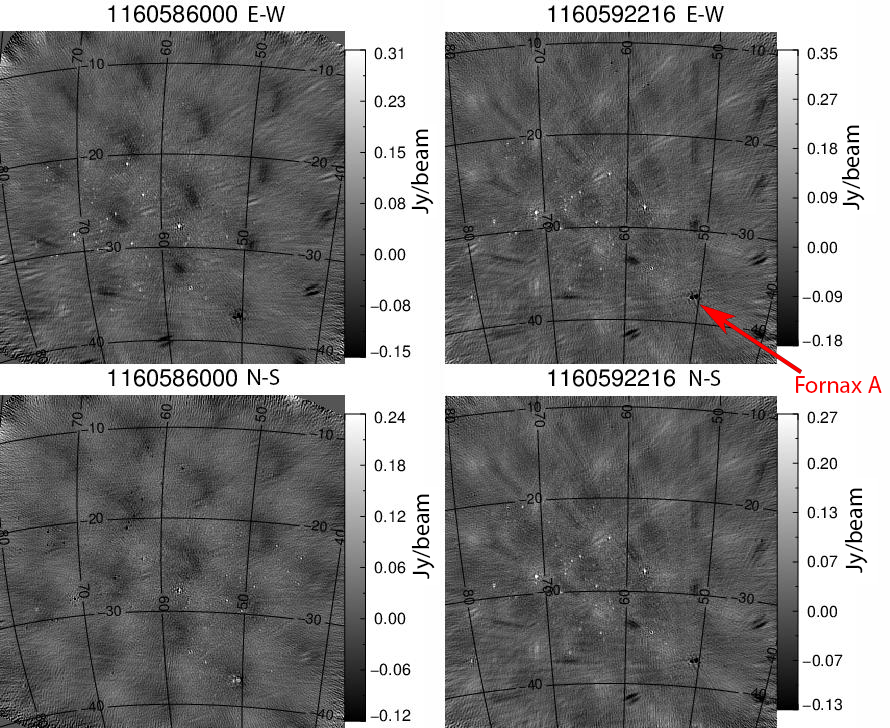}
    \caption{Residual images for both the E-W and N-S polarization of a zenith observation (observation id: 1160592216) and an off-zenith observation (observation id: 1160586000). Upper left: residual image for E-W off-zenith; Bottom left: residual image for N-S off-zenith; Upper right: residual image for E-W zenith; Bottom right: residual image for N-S zenith.}
    \label{fig:polarization}
\end{figure*}

In Section~\ref{sec:PS}, we saw that the PS improvements from redundant calibration in the EoR1 data were strongly polarization-dependent: for the E-W polarization, we saw substantive and consistent improvement at all pointings; for the N-S polarization, there were no improvements at off-zenith pointings and, while there was improvement at the zenith pointing, it was less than that of the E-W polarization. 

To explain the strong polarization dependence in redundant calibration's impact on the PS, we picked one zenith observation and one off-zenith observation of EoR1 night1 and investigated their FHD products. In particular, we looked at the residual images for both polarizations, which are generated from calibrated, residual visibilities, as shown in Figure~\ref{fig:polarization}. 
As mentioned in Section~\ref{sec:compare_eor0_eor1}, Fornax A is located near the centre of EoR1. Looking at residual images for both polarizations from an EoR1 zenith observation (observation ID 1160592216, right hand column), leads us to conclude that our Fornax A model contains too much flux density, leading to the negative hole at RA $50\deg$, Dec $-37\deg$.  This error clearly produces many other artifacts in the image, including a strong set of negative features associated with sidelobes of Fornax A and a general miscalibration and undersubtraction of the numerous other points sources in the field.

This pattern is seen in both polarizations of the zenith pointing---both of which are improved by redundant calibration.
We also see the over-subtraction of Fornax A and under-subtraction of other sources in the E-W residual image from an EoR1 off-zenith observation (observation ID 1160586000, left hand column). This is consistent with the fact that redundant calibration improves the E-W PS of an off-zenith observation. However, the N-S residual image looks quite different. Instead, we see an undersubtraction of Fornax A and oversubtraction of many point sources in the upper left of the image.  The sidelobes of Fornax A are also significantly mitigated.  

The most likely explanation for this effect is an error in our model of the tile primary beam. 
For any given pointing, the primary beam differs between the two linear polarizations.  At zenith, this is simply because the dipoles for detecting one polarization are rotated 90 degrees from each other (and the beam pattern of a dipole --- and therefore our tiles as well --- is not symmetric under a 90 degree rotation).  For pointings off-zenith, the difference between the two polarizations can be even more complicated, as the projection of the tile towards the pointing center can make the effective spacing between one polarization's dipoles appear foreshortened by a different amount than the other.  In general, our beam models are best for zenith, but become less accurate for other pointings.

The key impact for this work is that the beam model error made at the position of a specific source (e.g. Fornax A) will \emph{not} be the same in the two polarizations.
What we find is that, in the N-S off-zenith pointing, a fortuitous combination of error in the model of Fornax A itself and the polarization-dependent beam model error reduces the total level of sky-model error. In the E-W polarization and the N-S zenith-pointing the model error in Fornax A is the dominant source of error in our calibration solutions; such an error can be mitigated by redundant calibration, explaining
why we see PS improvements with {\tt \string OMNICAL} in all the observations except the N-S off-zenith pointings. This explanation is consistent with Figure~\ref{omni_solutions_EoR1_nights}, where the $|\Delta|$'s at the N-S polarization are smaller than those at the E-W polarization, particularly for off-zenith pointings.

\section{Conclusions}
\label{sec:conclusion}

In this paper, we studied the performance of tandem redundant and sky-based calibration in MWA Phase II data analysis in the EoR1 field. Using FHD and {\tt\string OMNICAL} in tandem, we calibrated and processed three nights of EoR1 observations, as well as one night of EoR for comparison with \cite{Wenyang_2019}. We performed a simulation to study redundant calibration's performance on fitting noise. We also analyzed calibration solutions and improvements in PS.  Redundant calibration solutions were found to have night-to-night consistency and resulted in repeatable, substantive improvements in the PS of the EoR1 field. We also saw the strong polarization dependence of redundant calibration's impact on the PS. Our principal findings are as follows:

\textbf{Redundant calibration gives non-negligible gain solutions.}
Our simulation shows that redundant calibration produces changes to the FHD sky-based gain solutions that are larger than would be expected if it were simply fitting to noise. 

\textbf{Redundant calibration's performance is repeatable.}
In our night-to-night comparisons, redundant calibration solutions and PS improvements are repeatable over the same range of LSTs. 

\textbf{Redundant calibration brings about substantive improvements in EoR1.} Generally, redundant calibration makes bigger improvements in the EoR1 PS than in EoR0 PS, including changes up to $10^4\  \text{mK}^2$ in the E-W polarization.

\textbf{Redundant calibration brings about larger improvements when the sky and beam model of FHD is worse.} Both the explanation for the bigger improvements in EoR1 PS than in EoR0 PS and our discussion for why we see the strong polarization dependence in redundant calibration's impact on PS tell us that when sky model incompleteness errors most affect sky-based calibration, redundant calibration can result in larger improvements in the PS.

These findings suggest that redundant calibration can continue to play an important role in 21\,cm cosmology analysis going foreward.  While the results from EoR0 suggest that with a good enough sky model for sky-based calibration, redundant calibration becomes ineffective, we note that the limits from \cite{barry_et_al_2019b}, \cite{Wenyang_2019}, and \cite{trott2020} are all still one to two orders of magnitude above any plausible EoR signal models. Therefore, all we can conclude is that the errors that can be mitigated by redundant calibration are not the dominant source of error in those works.  The present work shows that redundant calibration can indeed mitigate some level of sky-based calibration error and so it may continue to yield PS improvements once the current leading sources of systematic errors are mitigated.  To end on a note of caution, however, we should be prepared for the possibility that true non-redundancy in the array may become a real problem for redundant calibration and that, once lower PS levels are reached, the inclusion of redundant calibration algorithms may introduce more errors than they mitigate.  Research into new approaches that can balance sky-based and redundancy-based errors (e.g. \citealt{sievers2017} and \citealt{byrne2020}) is therefore of particular interest.

\section*{Acknowledgements}

ZZ, JCP, and WL would like to acknowledge the support from NSF grant \#1613040. BJH and MFM would like to acknowledge the support from NSF grants \#1506024, \#1613855, and \#1643011.
WL would like to acknowledge the Galkin Foundation Fellowship. 
KT is partially supported by Grand-in-Aid from the Ministry of Education, Culture, Sports, and Science and Technology (MEXT) of Japan, No. 15H05896, 16H05999 and 17H01110, and Bilateral Joint Research Projects of JSPS.
This scientific work makes use of the Murchison Radio-astronomy Observatory, operated by CSIRO. We acknowledge the Wajarri Yamatji people as the traditional owners of the Observatory site. Support for the operation of the MWA is provided by the Australian Government (NCRIS), under a contract to Curtin University administered by Astronomy Australia Limited. We acknowledge the Pawsey Supercomputing Centre which is supported by the Western Australian and Australian Governments. Parts of this research were supported by the Australian Research Council Centre of Excellence for All Sky Astrophysics in 3 Dimensions (ASTRO 3D), through project number CE170100013. CMT is supported by an ARC Future Fellowship under grant FT180100321. The International Centre for Radio Astronomy Research (ICRAR) is a Joint Venture of Curtin University and The University of Western Australia, funded by the Western Australian State government. The MWA Phase II upgrade project was supported by Australian Research Council LIEF grant LE160100031 and the Dunlap Institute for Astronomy and Astrophysics at the University of Toronto. This research was conducted using computation resources and services at the Center for Computation and Visualization, Brown University.

\bibliographystyle{pasa-mnras}
\bibliography{pasa-sample}

\end{document}